\documentclass[acmtog, authorversion]{acmart}
\acmSubmissionID{352}

\usepackage{booktabs} 
\usepackage{multirow}
\usepackage{caption}
\usepackage{subcaption}

\usepackage[ruled]{algorithm2e} 

\SetAlFnt{\small}
\SetAlCapFnt{\small}
\SetAlCapNameFnt{\small}
\SetAlCapHSkip{0pt}

\acmJournal{TOG}

\copyrightyear{2023}
\acmYear{2023}
\setcopyright{acmlicensed}\acmConference[SIGGRAPH '23 Conference Proceedings]{Special Interest Group on Computer Graphics and Interactive Techniques Conference Conference Proceedings}{August 6--10, 2023}{Los Angeles, CA, USA}
\acmBooktitle{Special Interest Group on Computer Graphics and Interactive Techniques Conference Conference Proceedings (SIGGRAPH '23 Conference Proceedings), August 6--10, 2023, Los Angeles, CA, USA}
\acmPrice{15.00}
\acmDOI{10.1145/3588432.3591504}
\acmISBN{979-8-4007-0159-7/23/08}

\begin{document}
\title{QuestEnvSim: Environment-Aware Simulated Motion Tracking from Sparse Sensors}

\makeatletter
\newcommand\addauthornote[1]{%
  \if@ACM@anonymous\else
    \g@addto@macro\addresses{\@addauthornotemark{#1}}%
  \fi}
\newcommand\@addauthornotemark[1]{\let\@tmpcnta\c@footnote
   \setcounter{footnote}{#1}\addtocounter{footnote}{-1}
    \g@addto@macro\@currentauthors{\footnotemark\relax\let\c@footnote\@tmpcnta}}
\makeatother

\author{Sunmin Lee}
\orcid{0009-0007-7482-807X}
\email{sunmin.lee@imo.snu.ac.kr}
\affiliation{%
 \institution{Seoul National University}
 \country{South Korea}
 }

\author{Sebastian Starke}
\orcid{0000-0002-4519-4326}
\email{Sebastian.Starke@mail.de}
\affiliation{%
 \institution{Reality Labs Research, Meta}
 \country{United States of America}
}

\author{Yuting Ye}
\orcid{0000-0003-2643-7457}
\email{yuting.ye@gmail.com}
\affiliation{%
 \institution{Reality Labs Research, Meta}
 \country{United States of America}
}

\author{Jungdam Won} 
\authornote{co-corresponding authors}
\orcid{0000-0001-5510-6425}
\email{jungdam@imo.snu.ac.kr}
\affiliation{%
 \institution{Seoul National University}
 \country{South Korea}
}

\author{Alexander Winkler} 
\authornotemark[1]
\orcid{0000-0003-1839-0855}
\email{alexander.w.winkler@gmail.com}
\affiliation{%
 \institution{Reality Labs Research, Meta}
 \country{United States of America}
}

\begin{abstract}
Replicating a user's pose from only wearable sensors is important for many AR/VR applications. Most existing methods for motion tracking avoid environment interaction apart from foot-floor contact due to their complex dynamics and hard constraints. However, in daily life people regularly interact with their environment, e.g. by sitting on a couch or leaning on a desk. Using Reinforcement Learning, we show that headset and controller pose, if combined with physics simulation and environment observations can generate realistic full-body poses even in highly constrained environments. The physics simulation automatically enforces the various constraints necessary for realistic poses, instead of manually specifying them as in many kinematic approaches. These hard constraints allow us to achieve high-quality interaction motions without typical artifacts such as penetration or contact sliding. We discuss three features, the environment representation, the contact reward and scene randomization, crucial to the performance of the method. We demonstrate the generality of the approach through various examples, such as sitting on chairs, a couch and boxes, stepping over boxes, rocking a chair and turning an office chair. We believe these are some of the highest-quality results achieved for motion tracking from sparse sensor with scene interaction.
\end{abstract}

%
%


%
%


\begin{teaserfigure}
    \centering \includegraphics[width=0.49\linewidth]{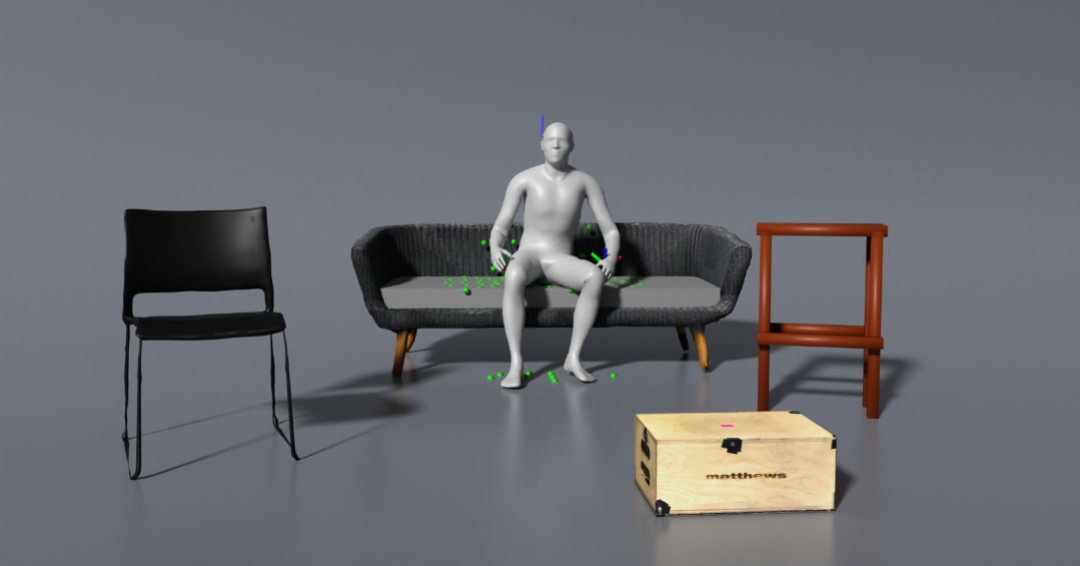}
    \includegraphics[width=0.49\linewidth]{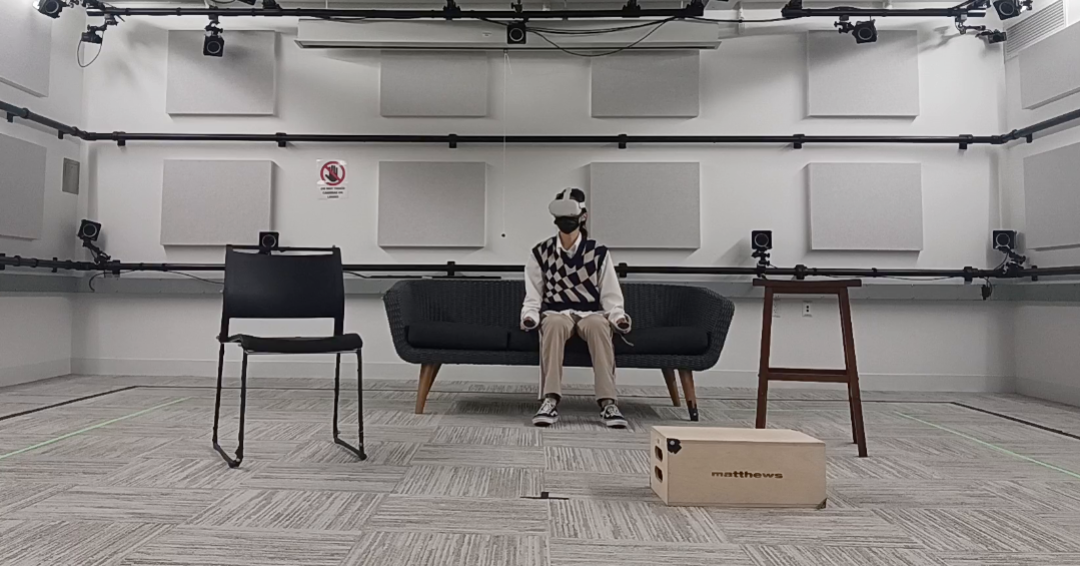}
    \caption{From only the pose of a headset and controllers, our method reconstructs a matching full-body pose that interacts naturally with various objects in a simulated environment.}
    \label{fig:teaser}
\end{teaserfigure}

\maketitle

\section{Introduction}

AR/VR (Augmented, Virtual Reality) has the potential to create entirely new social experiences. For example, instead of 2D video calls, users could interact in a virtual 3D space. To create the feeling of presence, a user's avatar must accurately replicate their movement and body language as well as naturally interact with the environment. 
Marker-based motion tracking is inconvenient for such an experience, since it requires dozens of expensive cameras, a special suit and tedious calibration procedures. 
To create a lower friction experience, recent works have investigated \textit{markerless} solutions for motion tracking such as phone video, Inertial-Measurement Units (IMUs), or Head Mounted Displays (HMDs).

We aim for creating a motion tracking system that allows environment interaction and relying only on the pose of the consumer VR device and environment information as input. 
Synthesizing full-body motions from sparse sensors is challenging because many different poses could potentially match a given sensor input and generation of the lower body motions is even more challenging due to the absence of information.
In addition to these challenges, generating plausible object-interaction motion requires special care. When users interact with their environments (e.g. sitting on a couch or leaning on a desk), the system should generate motions using the environments, which introduces complex physical constraints.
Also, such motions are different from locomotion since the lower body is not always fully constrained by balancing, so there is more ambiguity. For example, when sitting on a couch, many different poses could potentially match a given sensor input.

In this paper, we develop a motion tracking algorithm that takes as input the headset and controller pose as well as a representation of the environment and generates full-body motions that match both sensor inputs and its surrounding environment. More specifically, we use a physically simulated avatar and learn a control policy to generate torques to drive the simulated avatar via deep reinforcement learning, where the goal is to track the user's headset and controller pose as close as possible. Similar to our approach, several motion tracking systems using physics-based avatars have been proposed, however, environment interactions apart from foot-floor contact were not demonstrated~\cite{Winkler:2022:QuestSim, Ye:2022:Neural3Points}. Other methods ~\cite{Luo:2022:Embodied} incorporated artificial forces to deal with the complex contact dynamics, however these forces can produce unnatural motions. Instead of using artificial forces, our control policy is trained to actively use the environment to generate the appropriate external forces to drive the simulated avatar, where the strategy is learned from mocap data that includes environment interaction. As a result, our system generates motions that are physically accurate and more believable within their environment. For example, an HMD close to a chair likely implies a user has sit down compared to just being in a crouching position.

As contributions, we first demonstrate that sparse-upper body input, if combined with physics simulation and environment observations, can generate realistic full-body motions in highly constrained environments without using any artificial force. We also show a non-trivial combination of key technical components are crucial for achieving high tracking performance and generalization to unseen real user inputs: Environment representation fed into the policy, contact reward during learning control policies, and scene randomization. To show the capability of our system, various examples, such as sitting on chairs, a couch and boxes, stepping over boxes, rocking a chair and turning an office chair are demonstrated where all motions are generated from unseen real user inputs without using any cleanup or post-processing (e.g. inverse-kinematics, contact resolving, smoothing, and etc). We believe these are some of
the highest-quality results achieved for motion tracking from sparse sensor with scene interaction.
We also show the capability of a policy tracking users with scene interaction from the HMD alone without controllers and a policy without future observations. Finally, we run ablation studies for the key design choices adopted in our system to understand how they affect performance and generalization of our system.

\begin{figure*}[!htb]
    \includegraphics[width=0.9\linewidth]{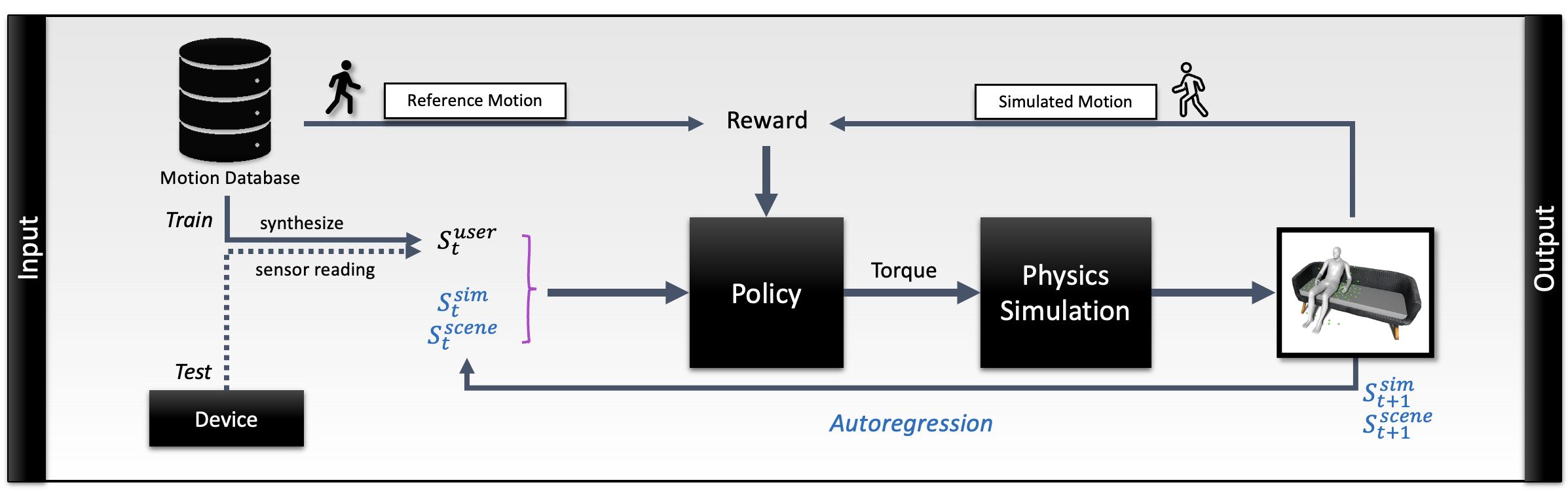}
    \caption{System Overview.}
    \label{fig:overview}
\end{figure*}

\section{Related Work}

Motion tracking/reconstruction from a specific sensor is a research topic that has a long history in computer graphics and computer vision, so we review previous studies that are most closely relevant to our approach, which includes \textit{synthesizing motions with interaction}, \textit{kinematic motion tracking from sparse sensor input}, and \textit{physics-based motion tracking}.

\subsection{Motion Synthesis with Interaction}

Synthesizing plausible full-body motions with interaction is regarded as one of the notorious problems because the difficulty grows exponentially as environments get more complex (e.g. the number of objects in the scene, the types of interactions). Several data-driven methods have been proposed for solving the problem.  Safonova et al.~\shortcite{Safonova:2007:Construction} optimized transitions in the pre-constructed motion graph where interactions were modeled as spatio-temporal constraints. Its extension to multi-character interaction was also demonstrated~\cite{Won:2014:Generating} via combining stochastic sampling and Laplacian motion editing~\cite{Kim:2009:Synchronized,Ho:2010:Spatial}.  Due to the complexity of optimization, patch-based methods have been proposed, which preserve interaction in a near-fixed state~\cite{Lee:2006:MotionPatches, Shum:2008:InteractionPatch, Henry:2012:Environment-Aware}. Although complex and large-scale human-object and human-human interactions have been demonstrated, they are computationally expensive and offline. Recently, many deep learning methods have been proposed, where a mapping from environment states to motions is constructed by a deep neural network in a supervised manner.  Holden et al.~\shortcite{Holden:2017:Phase-Functioned} demonstrated a locomotion controller autoregressively generating walking motions that adapt to uneven terrain. This idea was extended for human-scene interaction~\cite{Starke:2019:NSM} and hand-object interaction~\cite{Zhang:2021:ManipNet} where spatial representations such as voxel occupancy and proximity sensors were incorporated to describe the interactions. 
Our method is also a data-driven method and adopts spatial representations similar to the methods above to represent the current state of the environment, however, we synthesize such motions with sparse signals only and less supervision for the dataset.

\subsection{Motion Tracking from Sparse Sensors}

Using wearable sparse sensors for human motion tracking received a lot of attention due to its ease of use and wide applications in AR/VR (e.g. daily and outdoor mocap).  Several systems based on inertial measurement units (IMUs) have been proposed.  Marcard et al.~\cite{Marcard:2017:SIP} takes an offline approach optimizing pose parameters for the entire frame so that they match with the input signals.  The more popular approach is to use data-driven algorithms, where many proposed systems rely on a mocap database from which those systems search motion clips (or short segments) that match the input signals~\cite{Tautges:SIG:2011, Liu:2011:Realtime, Riaz:GM:2015, Andrews:CVMP:2016, Ponton:2022:Combining}. Deep neural networks (DNNs) have shown promising results on handling high-dimensional and large data. DNNs learn a mapping from the sparse input signals to full-body pose through supervised learning with paired data.  Real-time systems predicting local poses with negligible delay were demonstrated~\cite{Huang:2018:DIP, nagaraj2020rnn} and global root motions have been improved~\cite{ Yi:2021:TransPose, guzov2021human, jiang2022avatarposer} 
.  Other than using IMUs, a system based on wearable electromagnetic field sensors was also proposed~\cite{kaufmann2021empose}.  Because these methods learn complex mappings directly from data while considering only kinematics it is crucial to have well-prepared and sufficient mocap data for training. In it's absence, artifacts such as foot sliding or jittery motions can appear.  Furthermore, motion tracking from sparse sensors that allows full-body interactions has not been demonstrated yet in this research direction.

\subsection{Physics-based Motion Tracking}

Respecting physical laws when tracking user motions can improve the quality of generated motions. For example, foot sliding or penetration with the ground can be resolved by accurate contact modeling and forward simulation. One approach could be applying physics separately as soft constraints to refine estimated motions~\cite{Yi:2022:PIP}.  However, the most popular approach is to learn a control policy (a.k.a. controller) minimizing the discrepancy between the physically simulated characters and user inputs via deep reinforcement learning (Deep RL). Motion tracking from full-body mocap data have been proposed for a single clip~\cite{Merel:2017, Peng:2018:DeepMimic}, interactive kinematic motion controllers~\cite{Bergamin:2019:DReCon,Park:2019}, and large datasets~\cite{Chentanez:2018, won2020scalable, yuan2020residual, Levi:2021:SuperTrack}.  Similar systems using video clips as input (either ego-centric or 3rd-person view) have also been demonstrated~\cite{Peng:2018:SFV, Yuan:2019:Ego, Yuan:2021:SimPoE, Luo:2021:Dynamics}. They have also generated motions with contact-rich interaction such as sitting on chairs~\cite{Luo:2022:Embodied}.  We also generate motions with contact-rich interactions as demonstrated in~\cite{Luo:2022:Embodied}. However, our system only requires the pose of the headset and controllers as input, with no information about the lower body. We also don't use artificial forces to stabilize the simulated character.  

This setup is similar to recent approaches~\cite{Ye:2022:Neural3Points, Winkler:2022:QuestSim}. However, both approaches demonstrated only minimal environment interactions. In this paper we explicitly incorporate scene observations into the policy, which allows the policy to learn how to use the environment for motion tracking. The policy learns that sitting on a chair produces external forces that allows it to lift the leg, or expects an external force when stepping on a box raised above the floor. It also learns to manipulate external objects like a tilting chair through the appropriate contact forces. In general, this policy has a much better understanding of its environment and how external forces affect its pose.

\section{Method}

As input our method takes a sequence of poses (i.e. 6D transformations) from a user's VR headset and two hand controllers, where the user is interacting with daily-life objects (e.g. chairs, table, and etc). Our system generates a full-body motion that tracks the user and their environment interactions in a physically plausible manner. We assume the 3D scene geometry is known, and can be obtained by scanning the scene in advance. 
Figure~\ref{fig:overview} shows an overview of our method. 
We use a physically simulated avatar to enforce the naturalness in both generated motions and interactions. 
We train our policy using mocap data that includes environment interactions. 
Training dataset is explained in Section ~\ref{Training_Detail}.

We train a torque-based control policy to simulate the avatar via Deep Reinforcement Learning (DRL).
At each time step, an agent (i.e. the simulated avatar) observes the state $s_t$ and performs an action $a_t$. The state is updated to $s_{t+1}$ while receiving a reward $r_t$ that represents the desirability of the transition, the updated state, and the action. 
The goal of DRL is to learn a control policy $\pi_\theta(a_t|s_t)$ (i.e. a tracking controller represented as a deep neural network with parameters $\theta$), which maximizes the expected sum of rewards $J(\pi_\theta)=E[\sum_{t=1,...,\infty}{\gamma^{t-1}r_t}]$ over the entire trajectory, where $\gamma\in(0,1)$ is the discount factor. 
We treat the policy output $\pi(a_t|s_t)$ as the mean of a Gaussian distribution with a fixed, diagonal covariance matrix and use Proximal Policy Optimization (PPO) algorithm~\cite{DBLP:journals/corr/SchulmanWDRK17} to find an optimal policy. 

\subsection{Simulation Environment}
\begin{figure}
\includegraphics[width=0.46\columnwidth]{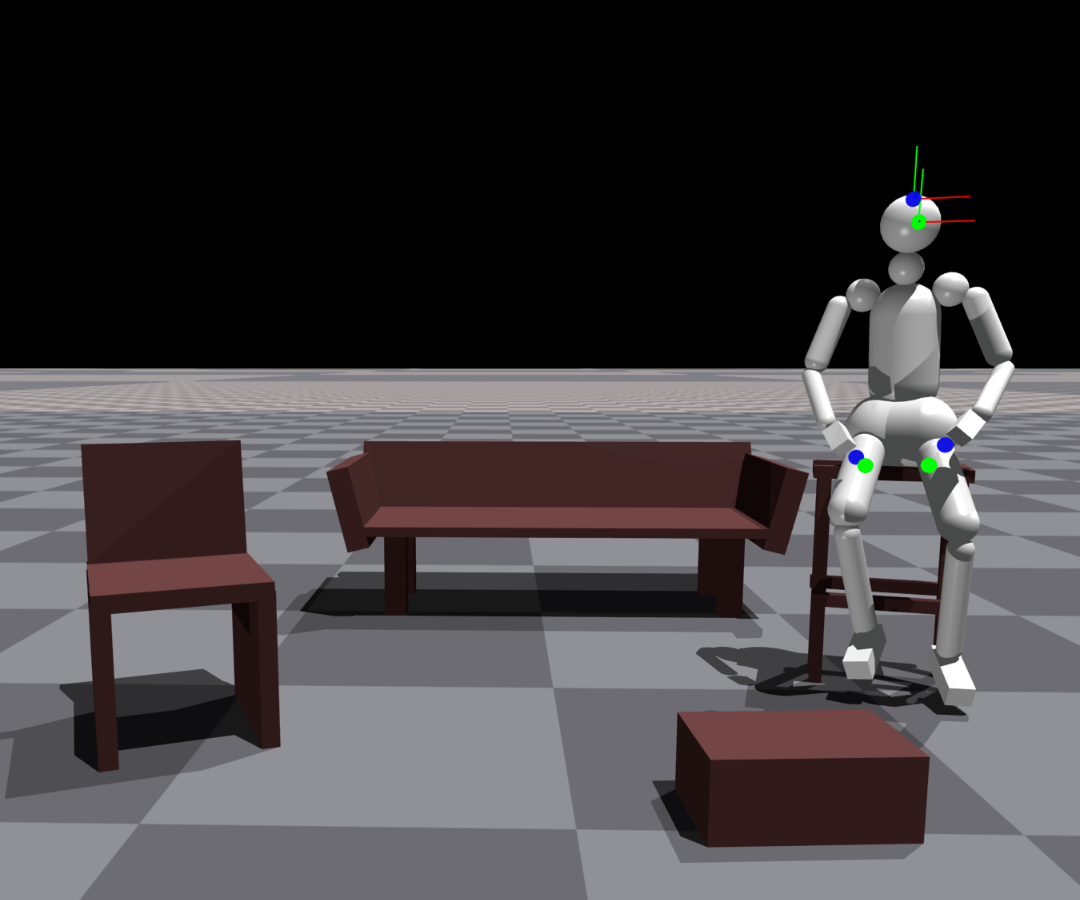}
\includegraphics[width=0.46\columnwidth]{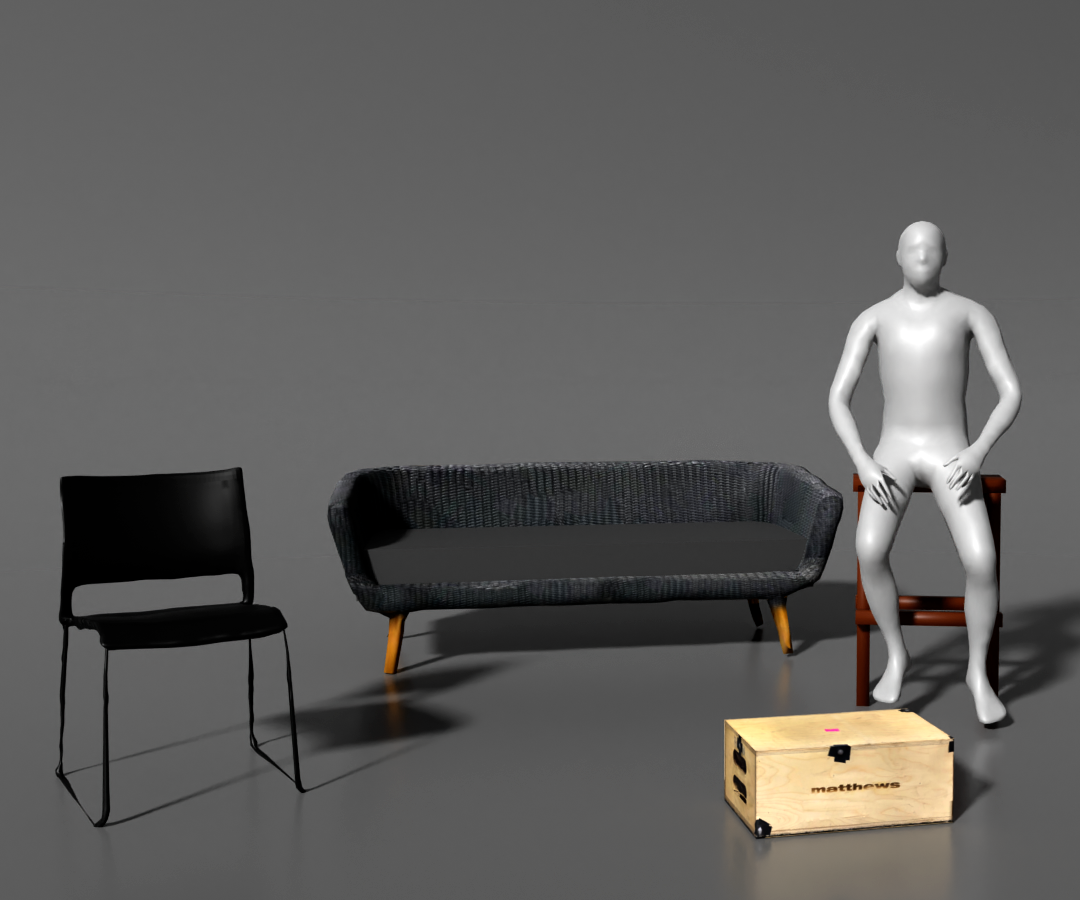}

\caption{For accurate contact behavior we approximate all the real shapes (right) with geometric primitives (left): The character's mesh is approximated with 18 collision geometries (capsules, boxes and spheres). The environment objects are similarly approximated with geometric primitives. As long as the contact surfaces match, the visualization will look accurate. We calculate masses and inertias of the character and objects based on their collision geometry and assuming uniform density.}
\label{fig:sim_vis}
\end{figure}

Our simulated character (Figure~\ref{fig:sim_vis}) has 32 degrees of freedom and 18 links and is driven by joint torques. 
We do not allow self-collision between the character's body links.
The objects that our mocap actors interact with must also be replicated in our physics simulation. We use primitive geometries like boxes or cylinders to approximate the real collision shapes. Figure~\ref{fig:sim_vis} shows an example scene that includes the simulated character and the environment objects. 

We label when and where contacts happen between the actor and their environment. This information is used as a supervision signal during training. The policy learns to leverage contacts with environment objects to track a user's motion. The contact labels are computed semi-automatically by using the simulator's collision detection algorithm on the mocap reference motions. Some manual correction is performed in difficult environments.

\subsection{State \& Action}
The policy outputs action in the range of [-1,1], which we scale to appropriate torques using each joints maximum torques. The state $s_t = (s^{sim}_t, s^{user}_t, s^{scene}_t)$ at time $t$ consists of the simulated avatar state $s^{sim}_t$, the pose of the users headset and controllers $s^{user}_t$ , and the scene observation $s^{scene}_t$ representing the surrounding environment and objects. These are discussed in the following.  

\subsubsection{Simulated State}
The simulated state 
$$s^{sim}_t = \{q, \dot{q}, p, \dot{p}, R, \dot{R}, f \}$$
is composed of joint angles $q$, link positions $p$, link orientations $R$, their corresponding velocities and contact forces $f$ acting on the links. All values are represented in the avatar-centric coordinate system (i.e. facing frame) as used in~\cite{won2020scalable, Winkler:2022:QuestSim}. This frame is defined by the pelvis orientation and position projected onto the ground. We use the first two columns of the rotation matrix to represent a link orientation.

\subsubsection{Sensor State}
The sensor state 
$$s^{user}_t=(o^{user}_{t-l},\cdots,o^{user}_{t},\cdots,o^{user}_{t+k})$$
is a time-window of user observations $o^{user}=(R_{h}, p_{h}, p_{l}, p_{r})$ which includes the orientation $R_{h}$ and the position $p_{h}$ of the headset and the positions $p_{l}$, $p_{r}$ from the left and right controllers. 

We only use the position of the controllers, not their orientation. This is because the controllers orientations proved to be noisy and less reliable, due to fast motion and user-dependent styles of holding controllers.
Similarly to the simulated character state, the sensor state is represented in the same avatar-centric coordinate system. In all our experiments, unless otherwise noted, we use a fixed window size of 1s of past and future information.

\subsubsection{Scene State}
In our method, information of the environment is crucial to make the policy understand and generate motions with environment interaction. The scene observation $s^{scene}_t = (h_1,h_2,\cdots) $
uses a height map to observe the geometrical features of the current environment. This height map is created from a circular grid centered around the avatar-centric coordinate and samples the height at each grid point $h_i$. 
The radius of the heightmap is 0.48m and it includes 120 grid points with 0.08m and 1/20 radian interval.

\subsection{Reward}
The behavior of the simulated avatar varies depending on the used rewards. In addition to an imitation reward that has been used in many other works~\cite{Peng:2018:DeepMimic, Bergamin:2019:DReCon, won2020scalable, Winkler:2022:QuestSim}, we also add rewards that are crucial for natural-looking motions with environment interaction. Our reward function 
\begin{equation}
r_t = r_\mathrm{imitation} + r_\mathrm{contact} + r_\mathrm{regularization}
\end{equation}
consists of three terms discussed in the following.

\subsubsection{Imitation Reward}
The imitation reward encourages to imitate the movement of the reference motion
\begin{equation}
\begin{aligned}
    r_{\text{imitation}} &= 
    w_q e^{-k_q \|q_\mathrm{sim}-q_\mathrm{ref}\|^2} + w_{\dot q} e^{-k_{\dot q} \|\dot{q}_\mathrm{sim}-\dot{q}_\mathrm{ref}\|^2} \\
    &+ w_p e^{-k_p \|p_\mathrm{sim}-p_\mathrm{ref}\|^2} + w_{\dot p} e^{-k_{\dot p} \|\dot{p}_\mathrm{sim}-\dot{p}_\mathrm{ref}\|^2} \\
    &+ w_R e^{-k_R \|log(R_\mathrm{sim},R_\mathrm{ref})\|^2},
\end{aligned}
\end{equation}
where we measure the difference between our simulated avatar and the reference motion by joint angles $q$, joint angle velocities $\dot q$, link positions $p$, link linear velocities $\dot p$, and link orientations $R$. The subscript $\mathrm{sim}$, $\mathrm{ref}$ refers to simulation and reference, respectively and $w_{q}$, $w_{\dot q}$, $w_{p}$, $w_{\dot p}$ and $w_{R}$ are the corresponding weights. In theory, this reward is over-specified because joints and links are simply transferable via forward kinematics. However, this over-specified formulation often performs better for many motion reconstruction tasks, not only in physics-based but also in kinematics-only settings~\cite{Peng:2018:DeepMimic, Holden:2017:Phase-Functioned, Starke:2019:NSM}. 

\subsubsection{Contact Reward}
The contact reward encourages the simulated avatar to create the same contact state as the reference character
\begin{equation}
    r_{\text{contact}} = w_{c} e^{-k_c \|c_\mathrm{sim}-\hat{c}_\mathrm{ref}\|},
\end{equation}
where the contact state $c=(c_{root}, c_{spine}, c_{feet}, c_{hands}$) is defined by multi-hot encoding which each element becomes 1 if the the link is currently in contact with any object in the scene, otherwise it becomes 0. We select feet, hands, pelvis and spine for the contact reward which are the links that take important roles in supporting the character's body and leave it open for other links.
During training, the contact state of the simulated character can be noisy, so we ignore contacts of magnitudes lower than $50N$ in the gravity (i.e. upward) direction. Note that this ground truth contact state is necessary only for training. In our all experiments, we observed that the absence of this contact reward significantly degrades overall motion quality (tracking performance and the naturalness). 

\subsubsection{Regularization Reward}
The regularization reward greatly improves the naturalness of the generated motions. 
\begin{equation}
    r_{\text{regularization}} = w_a e^{-k_a \|a_t\|^2} + w_s e^{-k_s \|a_t - a_{t-1}\|^2},
\end{equation}
where $a_t$, $a_{t-1}$ are the current and previous actions. Because the action in our simulation setup is a set of torques applied to all joints, the regularization reward prevents the controller from using excessive forces or changing forces abruptly.

\subsection{Scene Randomization}
We randomly vary the object placement during training to prevent the control policy from overfitting to the training data. 
Although larger object variation could further improve generalization, we randomized the scene in a small range so that the perturbed environment does not hugely contradict ground truth motion. It avoids the simulation becoming unstable during initialization due to the penetration and does not require motion editing, yet effective in generalization to unseen real-data In our experiment, the position of objects is randomly perturbed up to 8cm in all directions including height, and the orientation is randomly rotated up to 1 radian ($\approx$ 5.7 degrees) around the up-axis.

\section{Results}\label{Results}

\subsection{Training}\label{Training_Detail}
We implement our system using IsaacGym ~\cite{DBLP:journals/corr/abs-1810-05762} for physics simulation and Torch~\cite{NEURIPS2019_9015} for deep learning model. 
In every iteration of deep reinforcement learning, 61440 transition tuples are collected from 4096 simulated environments that are running in parallel on GPUs.
We use Proximal Policy Optimization (PPO) algorithm~\cite{DBLP:journals/corr/SchulmanWDRK17} where we use clip ratio 0.2, learning rate 1.2e-4, gamma 0.97, lambda 0.95, and minibatch size 7680, respectively. 
We use feed-forward deep neural networks with \textit{Tanh} activation for both actor (i.e. control policy) and critic (i.e. value function network), where the width and depth of those networks are $[300, 200, 100]$ and $[400, 400, 300, 200]$, respectively.

We create an in-house dataset (Table~\ref{table:Data_composition}) that includes motions such as sitting on various objects and stepping over boxes, as well as motions without interaction such as walking, gesturing, and squatting. 
We captured the dataset with 5 subjects, except for some motions (50 subjects for walking, gestures, and writing in the air). We modeled our character with a single body scale according to the subject's height.
We mirror all the data to double the data size.
We also captured object placement and motion if it moves.

To demonstrate various examples, we used a relevant subset of the total dataset to train the policy depending on the task.
We included non-interactive motions (walking, gestures, writing in the air) in all examples to enhance stable locomotion and hand tracking. 

\begin{table}[]
\caption{Training Data Composition}
\begin{tabular}{llll}
    \toprule
               & Time(Min) & Used Examples\\
   \midrule
Walking         & 140      & all       \\
Gestures & 140          & all \\
Writing in air & 116     & all      \\
Squat & 4.6         &   all \\
Bench Interaction   &      5.62    &     Living Room    \\
Couch Interaction   &  16.4       &  Living Room   \\
Sitting in Box, Chair, Stool  &  20.02 & Living Room \\ 
& & Tilting Chair \\
& & Rotating Chair \\ 
Tilting Chair & 6.98         &     Tilting Chair  \\
Rotating Chair & 3.27         &   Rotating Chair   \\
Stepping over Boxes   &    10.8        &     Stepping Box     \\
Sit on and get up from a floor  & 10.54       &   Getting up       \\
   \bottomrule
\end{tabular}
\label{table:Data_composition}
\end{table}

\begin{figure*}[!ht]
    \begin{subfigure}{0.495\linewidth}
        \centering
        \includegraphics[width=0.88\linewidth]{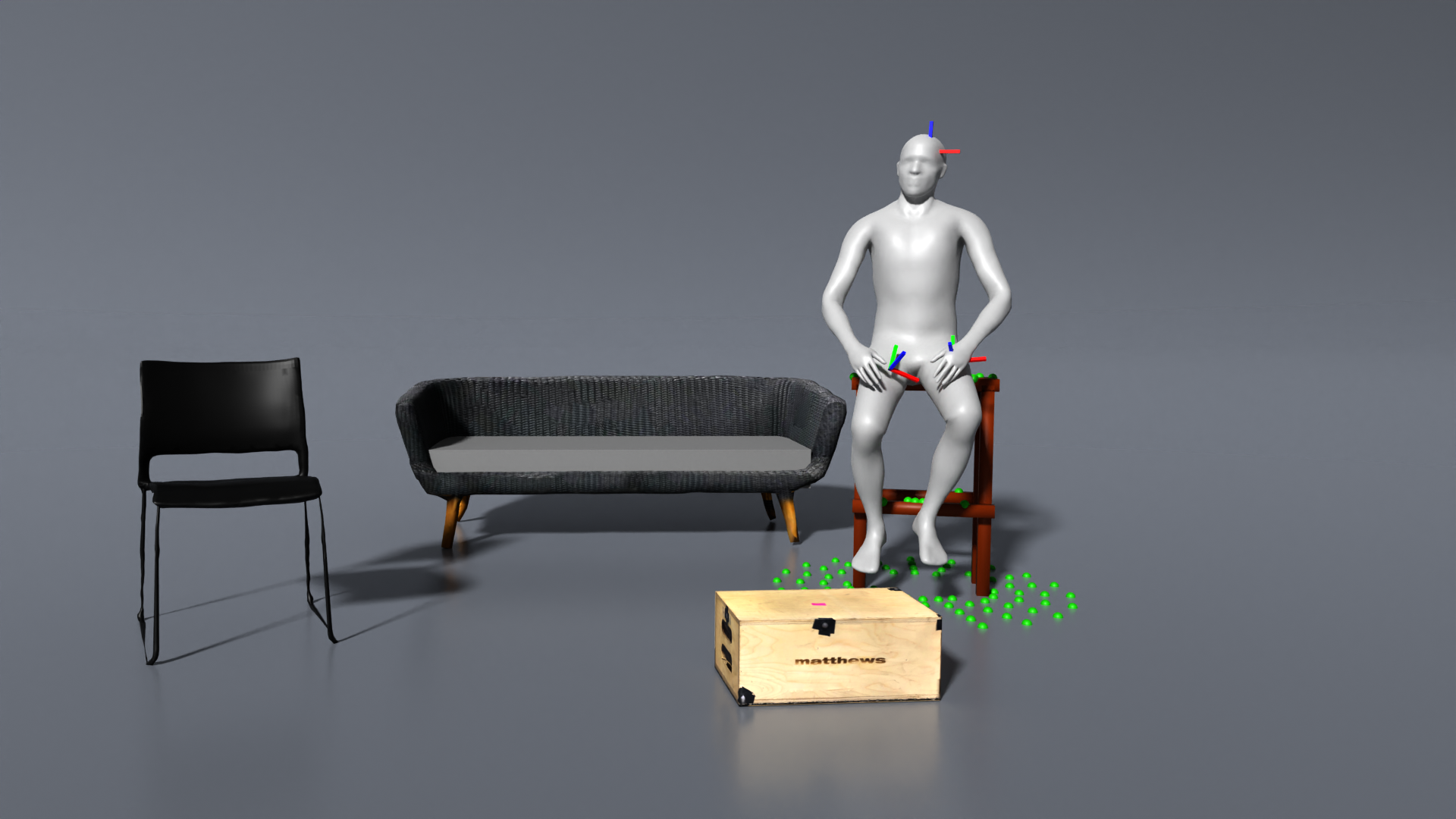}
        \caption{Living Room}
        \centering
        \includegraphics[width=0.44\linewidth]{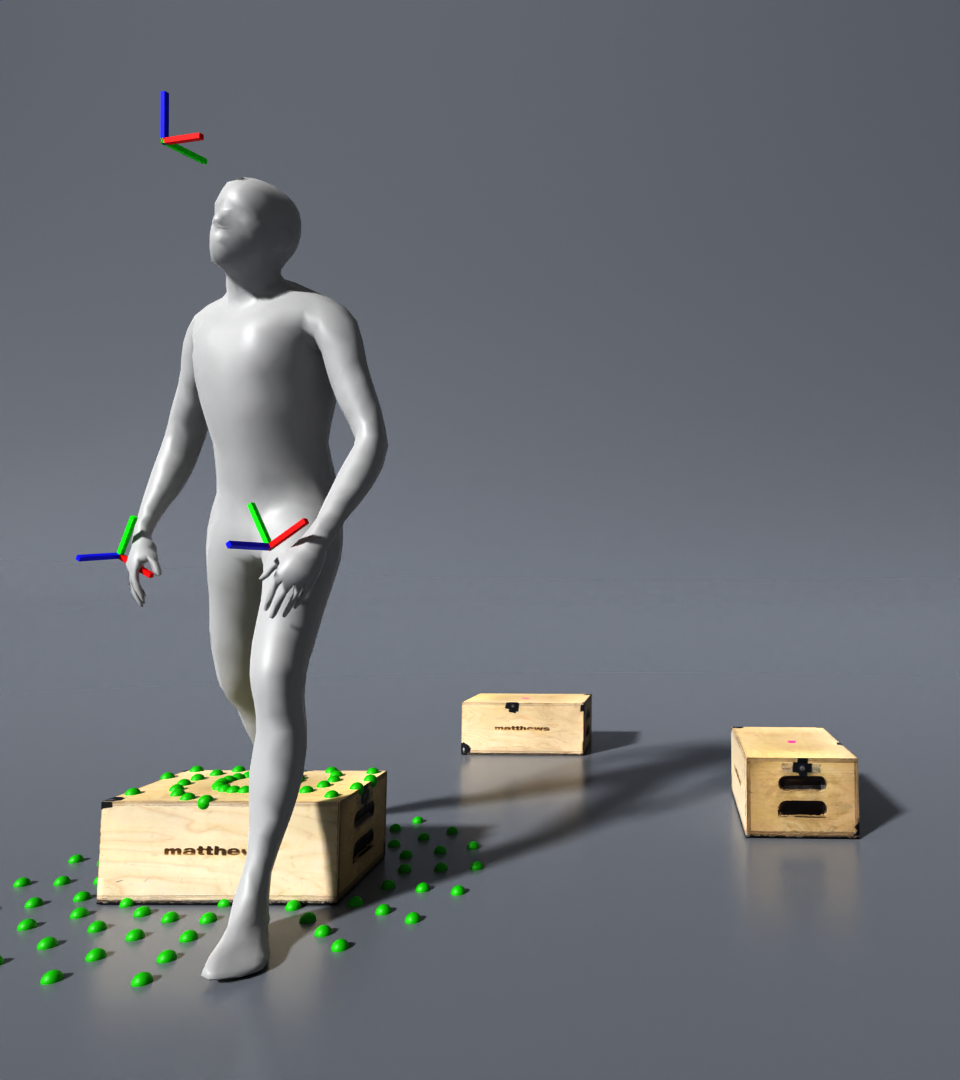}
        \includegraphics[width=0.44\linewidth]{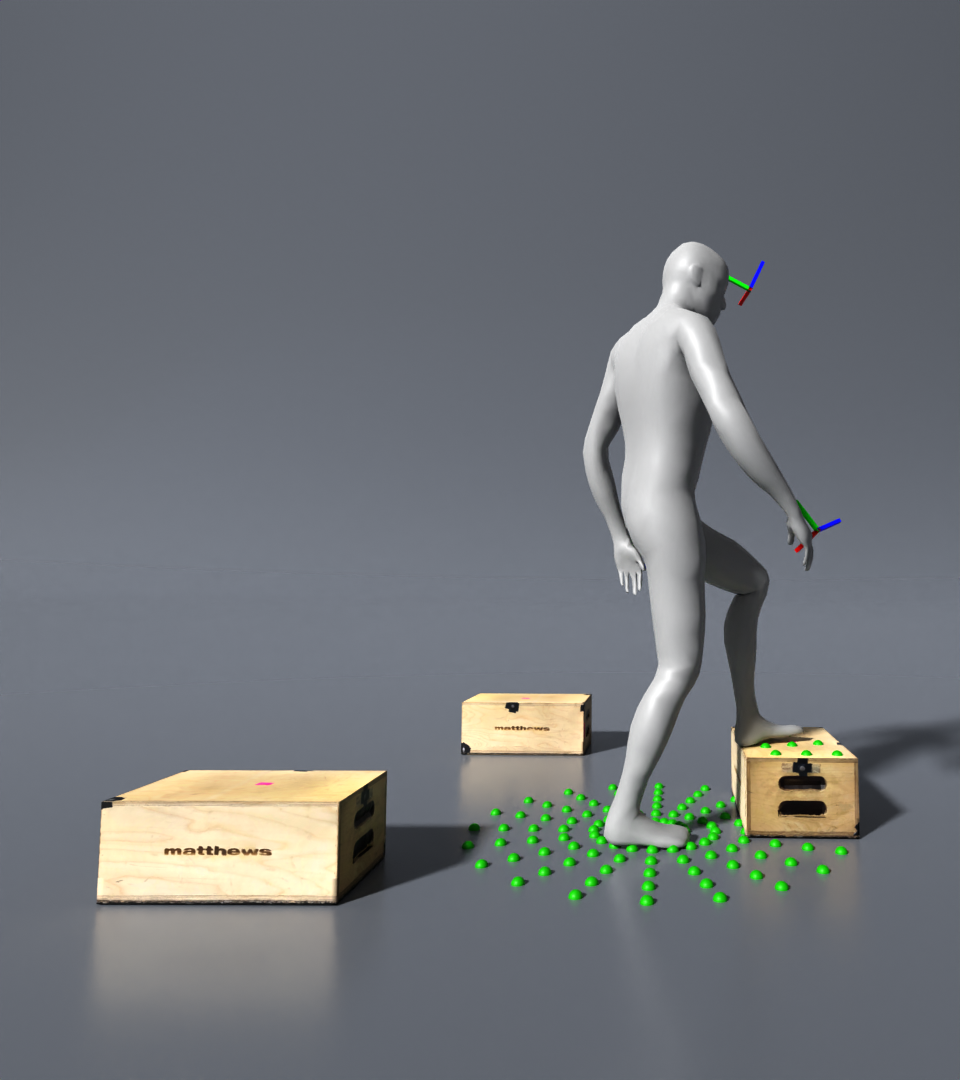}
        \caption{Stepping Box}
    \end{subfigure}
    \begin{subfigure}{0.495\linewidth}
        \centering
        \includegraphics[width=0.44\linewidth]{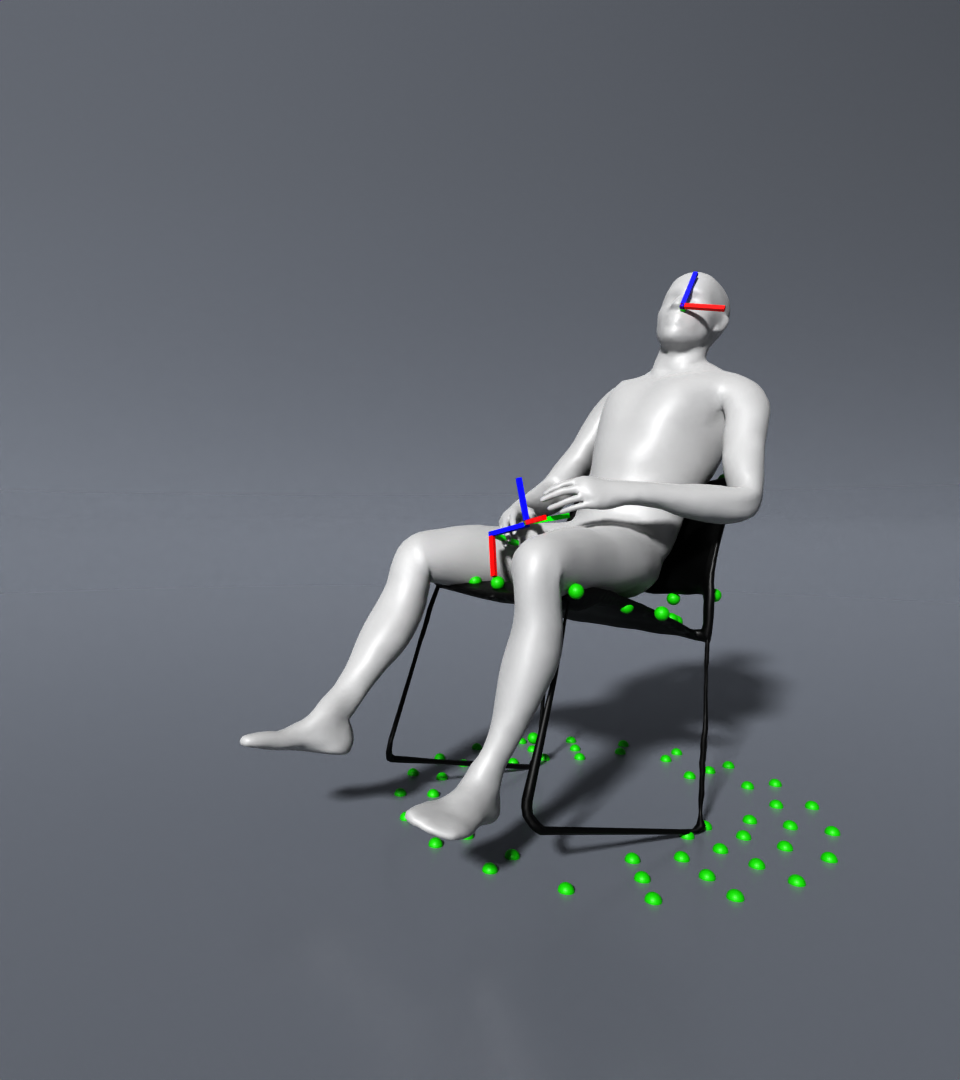}
        \includegraphics[width=0.44\linewidth]{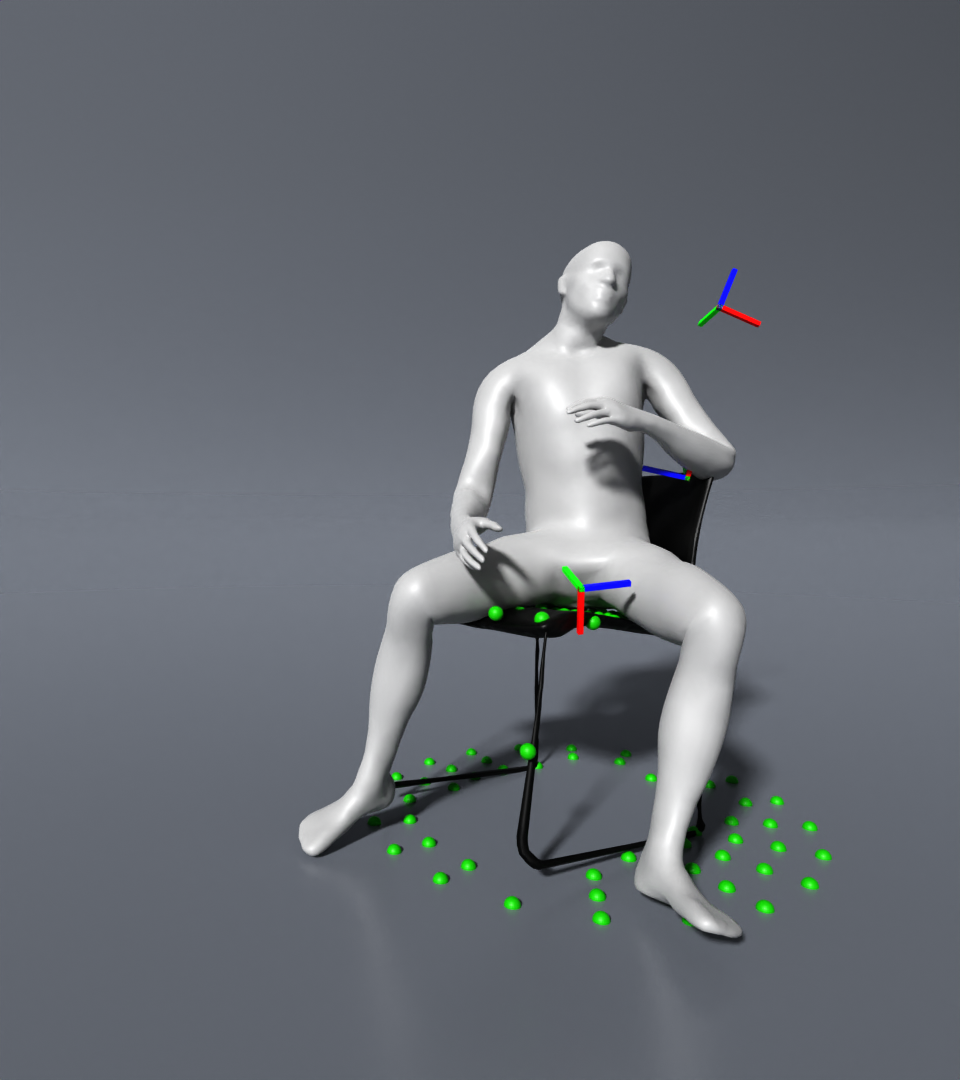}
        \caption{Tilting Chair}
        \centering
        \includegraphics[width=0.44\linewidth]{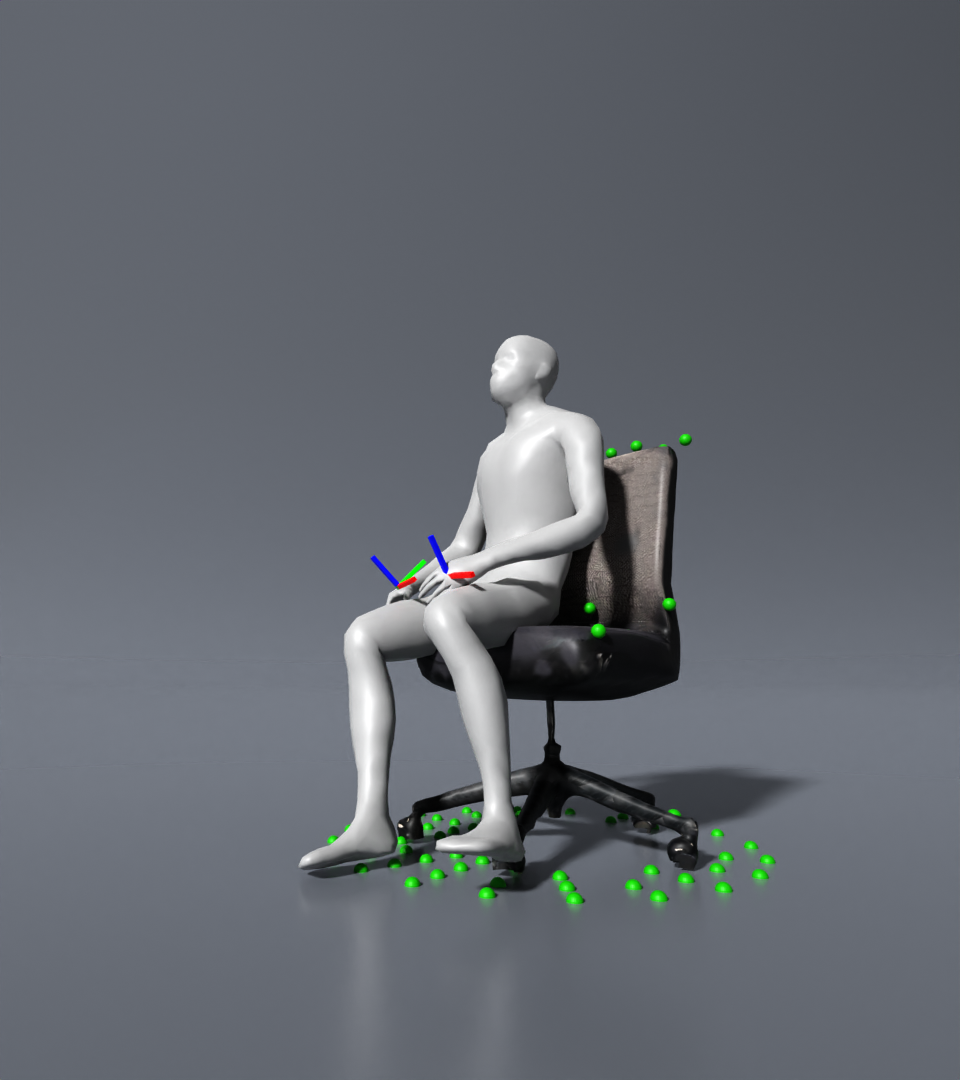}
        \includegraphics[width=0.44\linewidth]{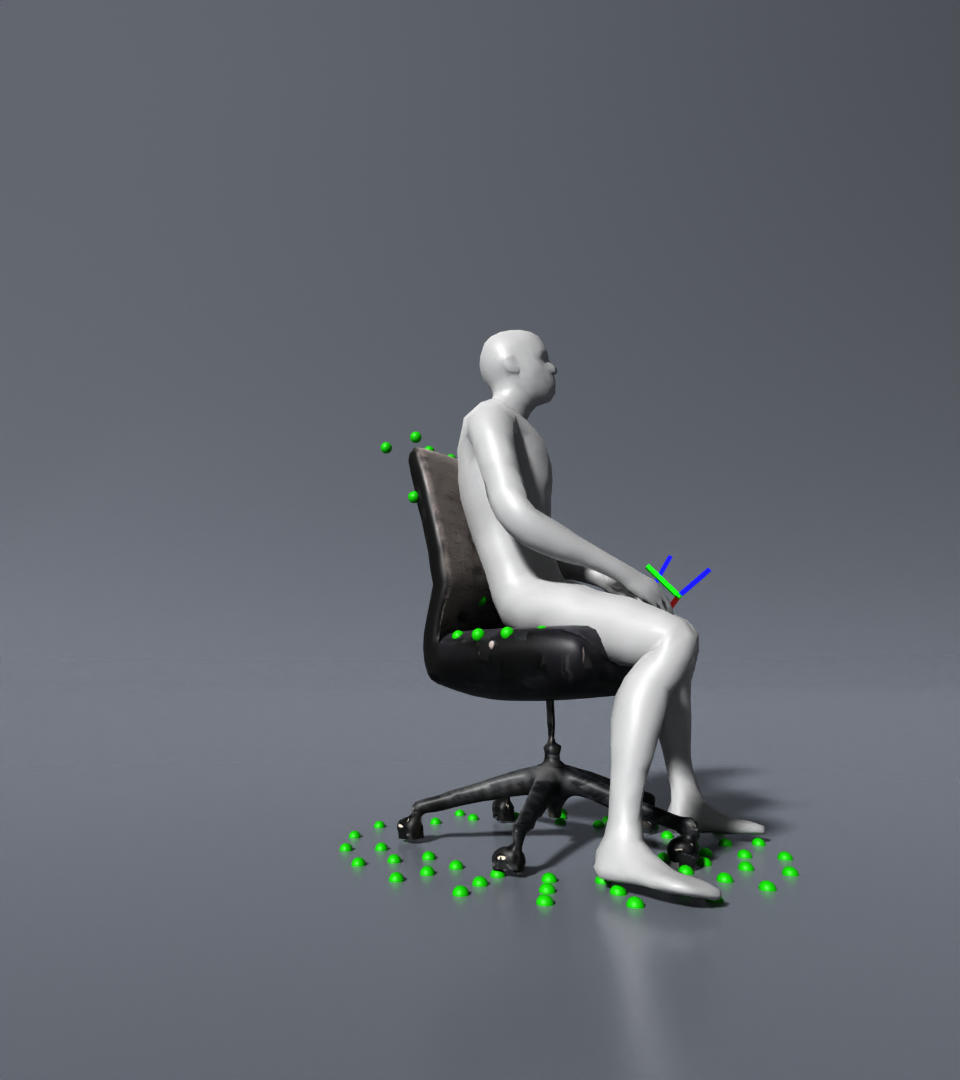}

        \caption{Rotating Chair}
    \end{subfigure}
    \caption{Examples Snapshot. The green spheres indicate height map around the simulated character.}
    \label{fig:example:default}
\end{figure*}

\begin{table*}[!h]
\caption{Performance for various scenarios using real sensors.}
\centering
\begin{tabular}{lccccccc} 
\hline
\multicolumn{1}{c}{~} & \begin{tabular}[c]{@{}c@{}}Tracking Error\\Headset [cm]\end{tabular} & \begin{tabular}[c]{@{}c@{}}Tracking Error\\Headset [deg]\end{tabular} & \begin{tabular}[c]{@{}c@{}}Tracking Error\\Controller [cm]\end{tabular} & \begin{tabular}[c]{@{}c@{}}Jerk \\{[}km/$s^3$]\end{tabular} & \begin{tabular}[c]{@{}c@{}}Success ratio\\{[}20s]\end{tabular} & \begin{tabular}[c]{@{}c@{}}Success ratio\\{[}30s]\end{tabular} & \begin{tabular}[c]{@{}c@{}}Success ratio\\{[}frame]\end{tabular} \\ 
\hline
Living Room & 5.6855 & 10.977 & 5.1963 & 0.3267 & 0.7689 & 0.6099 & 0.8239 \\
Stepping Box & 4.1150 & 7.1939 & 3.6289 & 0.4800 & 0.6683 & 0.4860 & 0.7422 \\
Tilting Chair & 6.4191 & 8.2919 & 5.3416 & 0.2303 & - & - & 0.7115 \\
Rotating Chair & 5.3879 & 14.0383 & 5.1379 & 0.2512 & - & - & 0.7623 \\
\hline
\end{tabular}
\label{table:performance:default}
\end{table*}

\begin{table*}[!h]
\caption{Impact of specific rewards, observations and training procedures on the living room example.}
\centering
\begin{tabular}{lccccccc} 
\hline
\multicolumn{1}{c}{~} & \begin{tabular}[c]{@{}c@{}}Tracking Error\\Headset [cm]\end{tabular} & \begin{tabular}[c]{@{}c@{}}Tracking Error\\Headset [deg]\end{tabular} & \begin{tabular}[c]{@{}c@{}}Tracking Error\\Controller [cm]\end{tabular} & \begin{tabular}[c]{@{}c@{}}Jerk\\{[}km/$s^3$]\end{tabular} & \begin{tabular}[c]{@{}c@{}}Success ratio\\{[}20s]\end{tabular} & \begin{tabular}[c]{@{}c@{}}Success ratio\\{[}30s]\end{tabular} & \begin{tabular}[c]{@{}c@{}}Success ratio\\{[}frame]\end{tabular} \\ 
\hline
Ours  & 5.6855 & 10.977 & 5.1963 & 0.3267 & \bf{0.7689} & 0.6099 & \bf{0.8239} \\
w/o Scene Observation & 5.7672 & 11.045 & 5.1422 & \bf{0.3017} & 0.73620 & \bf{0.6243} & 0.8185 \\
w/o Contact Reward & 5.6992 & \bf{9.8969} & 5.4907 & 0.4418 & 0.6060 & 0.3762 & 0.7147 \\
w/o Scene Randomization & \bf{5.5769} & 10.8448 & \bf{4.9798} & 0.4207 & 0.6010 & 0.4145 & 0.7202 \\
\hline
\end{tabular}
\label{table:performance:ablation}
\end{table*}

\begin{table*}[!h]
\caption{Impact of future sensor observations in the living room example. 3-point uses HMD and controllers, 1-point uses headset only.}
\centering
\begin{tabular}{lccccccc} 
\hline
\multicolumn{1}{c}{~} & \begin{tabular}[c]{@{}c@{}}Tracking Error\\Headset [cm]\end{tabular} & \begin{tabular}[c]{@{}c@{}}Tracking Error\\Headset [deg]\end{tabular} & \begin{tabular}[c]{@{}c@{}}Tracking Error\\Controller [cm]\end{tabular} & \begin{tabular}[c]{@{}c@{}}Jerk \\{[}km/$s^3$]\end{tabular} & \begin{tabular}[c]{@{}c@{}}Success ratio\\{[}20s]\end{tabular} & \begin{tabular}[c]{@{}c@{}}Success ratio\\{[}30s]\end{tabular} & \begin{tabular}[c]{@{}c@{}}Success ratio\\{[}frame]\end{tabular} \\ 
\hline
3-point with 1s future & \bf{5.6855} & \bf{10.977} & \bf{5.1963} & \bf{0.3267} & \bf{0.7689} & \bf{0.6099} & \bf{0.8239} \\
3-point with no future & 6.8037 & 12.3596 & 5.9104 & 0.4759 & 0.5948 & 
0.4845 & 0.7844 \\
1-point with 1s future & 6.1253 & 11.6988 & 12.1518 & 0.3527 & 0.6952 &  0.4001 & 0.7049 \\
\hline
\end{tabular}
\label{table:performance:robustness}
\end{table*}

\subsection{Evaluation}
To demonstrate the performance of our framework we show a variety of environment interactions, such as sitting on objects, getting up from the floor, manipulating objects as well as mismatches of real and virtual worlds that require environment observations. 
An overview of all the demonstrations can be seen in Figure ~\ref{fig:example:default}. We encourage readers to watch the accompanying video as it best demonstrates the physical reasoning our policy has learned and the quality of the motion tracking.

All evaluations were conducted in unseen object arrangement and user input. 
For instance, in the living room example, we combined a couch, a stool, and boxes that were in different training clips in different locations. We measured the object placement and replicated them in the simulation environment. We did not apply scene randomization during the evaluation.
Sensor inputs were directly fed from the real VR device. The simulated avatar is initialized in a default A-pose. A user does not have to start in a strict A-pose, and as long as they start standing, the system robustly catches up.

We also evaluate our system quantitatively by tracking error, jerk and success ratio in Table ~\ref{table:performance:default}.
When the average tracking error of the headset and controller is higher than 0.8m, we assume the character has drifted and consider it a failure. We measure the tracking error and jerk when successfully tracking the sensor and do not take them into account after the failure. We evaluate each test data multiple times by initializing in all possible frames. 
Success ratio [20s/30s] is computed as the ratio of number of episodes that did not fail for 20s and 30s. 
Success ratio [frame] indicates the average number of frames successfully tracked during 30s.

\subsection{Deliberate environment contact}
Most existing approaches are trained only on flat-ground and foot contact.
Our framework is able to deliberately generate contact forces with the environment to influence its root motion to track the sensors. 
One example is the getting up from the ground, where the policy uses the hands to generate appropriate contact forces.  

\begin{figure*}
    \includegraphics[width=0.87\linewidth]{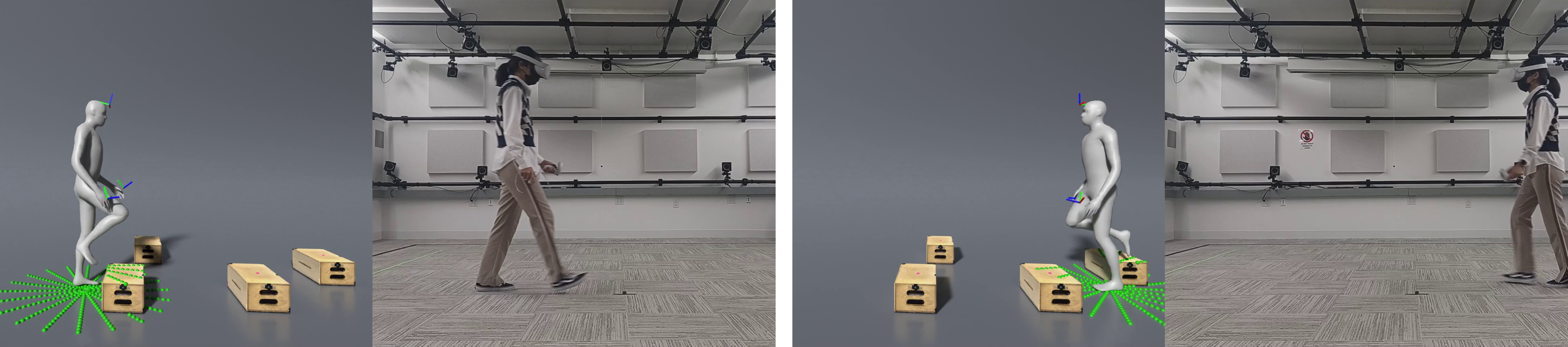}
    \caption{The user walks on flat ground, whereas we place the virtual boxes in the simulation environment. The policy learns to observe its surrounding scene by height map and to lift the leg higher to avoid obstacles while tracking the user's sensor data.}
    \label{fig:example:virtual_box}
\end{figure*}

The box-stepping examples demonstrate how the policy learned to step on a box to achieve a higher head position to track the real user. If there were no simulated box, the policy would not attempt the step. On the other hand our policy also learns to avoid environment contact if this contact is expected to interfere with the tracking (Figure ~\ref{fig:example:virtual_box}). We show a user walking on flat ground, whereas the simulation contains boxes at arbitrary positions. Since contact with these boxes will likely decrease headset and controller tracking, the policy learns to lift the leg higher than usual to avoid them.

\subsection{Manipulating Objects to improve tracking}
Tilting or rotating a chair is a very common behavior people do in daily life but it involves complex dynamics. With our simulated approach and environment observations, our approach learns to manipulate objects in a way to best track a user's motion.

We model the revolving chair with a revolute joint between the seat and the bottom. Our training data also includes moving objects such as the office chair, so the character learns that using feet to generate a frictional contact force with the ground creates a torque on the root. This torque can be used to rotate the chair in order to better follow the users HMD and controller positions. For the tilting chair, the character understands that pushing its feet into the ground also exerts a force on the backrest of the chair causing it to tilt. 
And it learned that tilting the chair in such a situation is beneficial for more closely tracking the user's sensor pose.

\subsection{Ablation on design choices}
Three crucial components are required to achieve the demonstrated performance: The scene observation in an appropriate representation, the contact reward during training which includes not just feet but various other body parts, and finally the random variation of the position of objects during training. In Table \ref{table:performance:ablation} we quantitatively show how each component improves the performance. We notice that especially leaving out the contact reward significantly reduces the success ratio [30s]. A similar drop in performance can be seen without scene randomization. Scene observations slightly increase the success ratio [20s] in this example. 
For this sitting example, the policy might be able to infer sitting from the quest pose alone. The scene observations show more usefulness in the box stepping videos, where the character raises its feet to avoid contact with a virtual box. 
Finally, even though some design choices don't significantly impact the tracking errors, the success corresponds more to perceived visual quality and is therefore prioritized.

\subsection{Ablation Headset only vs Headset and Controllers}
\begin{figure*}
    \includegraphics[width=0.9\linewidth]{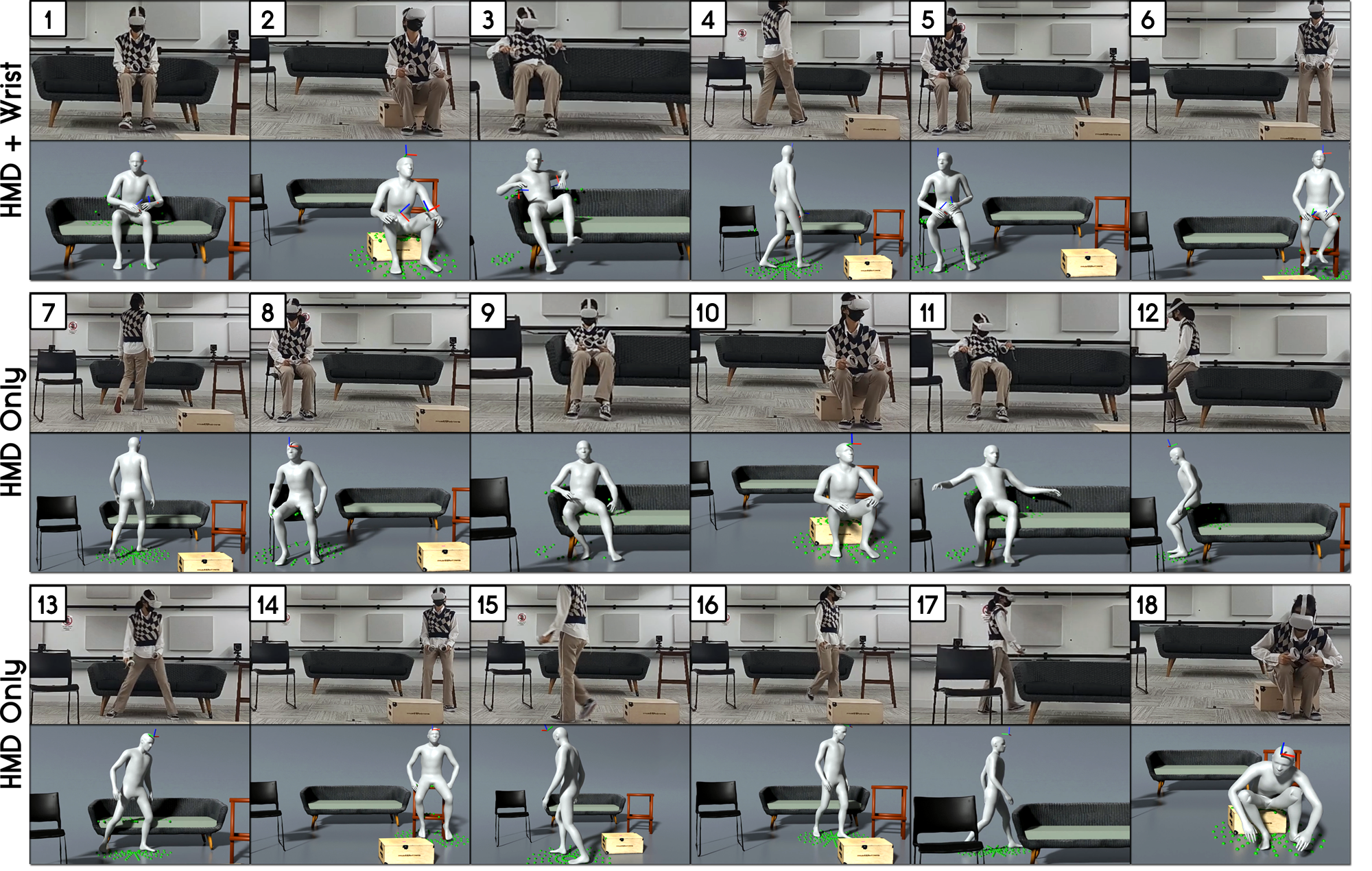}
    \caption{Collection of real user poses and simulated avatar poses using HMD and wrist trackers (top row) and only HMD tracker (middle and bottom rows).}
    \label{fig:collection}
\end{figure*}

We also demonstrate tracking users with scene interaction from the HMD alone, without controllers (Figure ~\ref{fig:collection} middle and bottom rows). Even with such limited input, the character is still able to interact with various objects, albeit with slightly worse metrics (Table ~\ref{table:performance:robustness} bottom row). 
In some situations, the model utilizes the arms in different ways than the user, such as when sitting on a chair (Figure ~\ref{fig:collection} (14)), in order to better stabilize the body. Due to the ambiguity of poses, the character may also produce different upper-body poses (Figure ~\ref{fig:collection} (11)), e.g. where the arms are hovering over the sofa while the real user is resting the arms on it.

\subsection{Ablation real-time vs future}
Our framework can be used to track users in real-time, as is shown in the video. However, the quality of the generated poses degrades compared to if the policy has access to one future observation. This degradation is reflected in Table \ref{table:performance:robustness} in the lower success ratio, as well as higher tracking errors. Nonetheless, as can be seen in the video, qualitatively the motions look reasonable, and the learned environment interaction of sitting and getting up is still possible. Access to future information will likely provide the biggest benefit for motions that require future planning, like jumping over a gap. However, to track day-to-day movements, real-time tracking provides acceptable results.

\section{Conclusion and Future Work}

We demonstrated a physics-based motion tracking framework from sparse sensors that actively utilizes physical interactions with the environment to generate natural-looking motions. The policies were trained using Reinforcement Learning, where our non-trivial combination of the components in the system design (e.g. environment observation, contact rewards, and scene randomization) enabled the policy to learn accurate and robust control strategies. We believe these are some of the highest-quality results achieved for motion tracking from sparse sensors with scene interaction. 

Although we showed the solid performance of our motion system over a variety of scenarios, there exist several limitations that we want to address in the future.
First, each type of interaction requires a specialized motion tracker in our current system. It would be ideal if a single tracker can be learned, which covers wider repertoires. 
This might need a more complex neural network model such as a mixture-of-experts~\cite{won2020scalable} or longer training time with larger datasets in general.

We also had a difficulty in reliably producing motions like getting up from the floor or generally more complex interaction motions. Since we don't use any artificial root forces, such behaviors that require careful coordination of contacts still seem difficult to learn (getting up from the floor is also a difficult motion to perform for older people). 
The simulated avatar used in our system can also fail (i.e. losing balance) as other physics-based characters, which incurs the failure of motion tracking of the user. Automatic failure detection followed by applying external force can be a reasonable compromise, however, physical realism would be degraded.

Another promising future direction would be extending our system for unknown scenes that include dynamically moving objects. In this case, the ability to infer physical properties (e.g. inertia, fiction coefficients) of surrounding objects in a scene would become more critical for successful motion tracking. Online system identification~\cite{yu2017preparing, feng2022genloco} could be combined as a part of the system.

\begin{acks}
We thank Nicky (Sijia) He for making 3D scans of environment objects and post-processing mocap data. We thank Michelle Hill and Tiffany Whitely for directing the motion capture sessions for data acquisition. We thank Jeongseok Lee for the integrated development support.
Jungdam Won and Sunmin Lee were partially supported by the New Faculty Startup Fund from Seoul National University and ICT(Institute of Computer Technology) at Seoul National University.
\end{acks}

\bibliographystyle{ACM-Reference-Format}
\bibliography{reference}

\newpage


\end{document}